\documentclass[final,times,5p]{elsarticle}
\usepackage{graphicx}
\usepackage{xspace}
\usepackage{amsmath}
\usepackage{amssymb}
\usepackage{amsfonts}
\usepackage{hyperref}

\newcommand{\GBll}{\ensuremath{\Gamma_{\ell\ell}\times\Gamma_{\ell\ell}/\,\Gamma}\xspace}
\newcommand{\GBee}{\ensuremath{\Gamma_{ee}\times\Gamma_{ee}/\,\Gamma}\xspace}
\newcommand{\GBmumu}{\ensuremath{\Gamma_{ee}\times\Gamma_{\mu\mu}/\,\Gamma}\xspace}
\renewcommand{\Re}{\ensuremath{\text{Re}}\,}
\renewcommand{\Im}{\ensuremath{\text{Im}}\,}
\renewcommand{\epsilon}{\varepsilon}

\journal{Physics Letters B}

\begin{document}

\begin{frontmatter}
  
  \title{Measurement of  
    $\Gamma_{ee}(J/\psi)\cdot\mathcal{B}(J/\psi\to e^+e^-)$ and 
    $\Gamma_{ee}(J/\psi)\cdot\mathcal{B}(J/\psi\to \mu^+\mu^-)$}

  

\author[binp]{V.V.~Anashin}
\author[binp,nsu]{V.M.~Aulchenko}
\author[binp,nsu]{E.M.~Baldin\corref{cor}} 
\cortext[cor]{Corresponding author, e-mail:  E.M.Baldin@inp.nsk.su}
\author[binp]{A.K.~Barladyan}
\author[binp]{A.Yu.~Barnyakov}
\author[binp]{M.Yu.~Barnyakov}
\author[binp,nsu]{S.E.~Baru}
\author[binp]{I.V.~Bedny} 
\author[binp,nsu]{O.L.~Beloborodova}
\author[binp]{A.E.~Blinov}
\author[binp,nstu]{V.E.~Blinov}
\author[binp]{A.V.~Bobrov}
\author[binp]{V.S.~Bobrovnikov}
\author[binp,nsu]{A.V.~Bogomyagkov}
\author[binp,nsu]{A.E.~Bondar}
\author[binp]{D.V.~Bondarev}
\author[binp]{A.R.~Buzykaev}
\author[binp,nsu]{S.I.~Eidelman}
\author[binp]{Yu.M.~Glukhovchenko}
\author[binp]{V.V.~Gulevich}
\author[binp]{D.V.~Gusev}
\author[binp]{S.E.~Karnaev}
\author[binp]{G.V.~Karpov}
\author[binp]{S.V.~Karpov}
\author[binp,nsu]{T.A.~Kharlamova}
\author[binp]{V.A.~Kiselev}
\author[binp,nsu]{S.A.~Kononov}
\author[binp]{K.Yu.~Kotov}
\author[binp,nsu]{E.A.~Kravchenko}
\author[binp,nsu]{V.F.~Kulikov}
\author[binp,nstu]{G.Ya.~Kurkin}
\author[binp,nsu]{E.A.~Kuper}
\author[binp,nstu]{E.B.~Levichev}
\author[binp]{D.A.~Maksimov}
\author[binp]{V.M.~Malyshev}
\author[binp]{A.L.~Maslennikov}
\author[binp,nsu]{A.S.~Medvedko}
\author[binp,nsu]{O.I.~Meshkov}
\author[binp]{S.I.~Mishnev}
\author[binp,nsu]{I.I.~Morozov}
\author[binp,nsu]{N.Yu.~Muchnoi}
\author[binp]{V.V.~Neufeld}
\author[binp]{S.A.~Nikitin}
\author[binp,nsu]{I.B.~Nikolaev}
\author[binp]{I.N.~Okunev}
\author[binp,nstu]{A.P.~Onuchin}
\author[binp]{S.B.~Oreshkin}
\author[binp,nsu]{I.O.~Orlov}
\author[binp]{A.A.~Osipov}
\author[binp]{S.V.~Peleganchuk}
\author[binp,nstu]{S.G.~Pivovarov}
\author[binp]{P.A.~Piminov}
\author[binp]{V.V.~Petrov}
\author[binp]{A.O.~Poluektov}
\author[binp]{I.N.~Popkov}
\author[binp]{V.G.~Prisekin}
\author[binp]{A.A.~Ruban}
\author[binp]{V.K.~Sandyrev}
\author[binp]{G.A.~Savinov}
\author[binp]{A.G.~Shamov}
\author[binp]{D.N.~Shatilov}
\author[binp,nsu]{B.A.~Shwartz}
\author[binp]{E.A.~Simonov}
\author[binp]{S.V.~Sinyatkin}
\author[binp,nsu]{Yu.I.~Skovpen}
\author[binp]{A.N.~Skrinsky}
\author[binp,nsu]{V.V.~Smaluk}
\author[binp]{A.V.~Sokolov}
\author[binp]{A.M.~Sukharev}
\author[binp,nsu]{E.V.~Starostina}
\author[binp,nsu]{A.A.~Talyshev}
\author[binp]{V.A.~Tayursky}
\author[binp,nsu]{V.I.~Telnov}
\author[binp,nsu]{Yu.A.~Tikhonov}
\author[binp,nsu]{K.Yu.~Todyshev}
\author[binp]{G.M.~Tumaikin}
\author[binp]{Yu.V.~Usov}
\author[binp]{A.I.~Vorobiov}
\author[binp]{A.N.~Yushkov}
\author[binp]{V.N.~Zhilich}
\author[binp,nsu]{V.V.~Zhulanov}
\author[binp,nsu]{A.N.~Zhuravlev}

  \address[binp]{Budker Institute of Nuclear Physics, 11, akademika
  Lavrentieva prospect,  Novosibirsk, 630090, Russia}
  \address[nsu]{Novosibirsk State University, 2, Pirogova street,  Novosibirsk, 630090, Russia}
  \address[nstu]{Novosibirsk State Technical University, 20, Karl Marx
  prospect,  Novosibirsk, 630092, Russia}

  \begin{abstract}

  The products of the electron width of the \(J/\psi\) meson and the
  branching fraction of its decays to the lepton pairs were
  measured  using data from the KEDR experiment at the VEPP-4M
  electron-positron collider. The results are 
\begin{equation*}
  \begin{split}
    &\GBee\,=0.3323\pm0.0064\,\text{(stat.)}\,\pm0.0048\,\text{(syst.)}\,\,\text{keV},\\ 
    &\GBmumu=0.3318\pm0.0052\,\text{(stat.)}\,\pm0.0063\,\text{(syst.)}\,\,\text{keV}.   
  \end{split}
\end{equation*}
Their combinations 
\begin{equation*}
  \begin{split}
    \Gamma_{ee}\times(\Gamma_{ee}+\Gamma_{\mu\mu})/\,\Gamma&=
    0.6641\pm0.0082\,\text{(stat.)}\,\pm0.0100\,\text{(syst.)}\,\text{keV,} \\
    \Gamma_{ee}/\,\Gamma_{\mu\mu}&=1.002\pm0.021\,\text{(stat.)}\,\pm0.013\,\text{(syst.)} \\
  \end{split}
\end{equation*}
can be used to improve the accuracy of the leptonic and full widths and test 
leptonic universality. 

Assuming $e\mu$ universality and using the world average value of the lepton
branching fraction, we also determine the leptonic
\(\Gamma_{\ell\ell}=5.59\pm0.12\,\text{keV}\) and total
\(\Gamma=94.1\pm2.7\,\text{keV}\) widths of the \(J/\psi\) meson.

  \end{abstract}

\end{frontmatter}

\section{Introduction}
\label{sec:intro}

The \(J/\psi\) meson is frequently referred to as a hydrogen atom for
QCD.  The electron widths \(\Gamma_{ee}\) of charmonium states are
rather well predicted by potential
models~\cite{Badalian:2008bi,lakhina-2006-74}. The accuracy in the
QCD lattice calculations of \(\Gamma_{ee}\) gradually approaches the
experimental errors~\cite{dudek-2006-73}.  The total and
leptonic widths of a hadronic resonance, \(\Gamma\) and
\(\Gamma_{\ell\ell}\), describe fundamental properties of the strong
potential~\cite{brambilla-2005}.

In this paper we report a measurement of the product of the electron 
width and the branching fraction to an \(e^+e^-\) pair for the 
\(J/\psi\) meson, \GBee.  An experimental
determination of \GBee requires scanning the beam energy and measuring the
cross section.  
In contrast to a measurement of the leptonic width itself, in this
case knowledge of the efficiency for hadronic decays does not
contribute to the final uncertainty.  
 The problem
considered can be reduced to measuring the area under the resonance
curve for the process \(e^+e^-\to J/\psi\to e^+e^-\).  Additionally we
have measured the product of the electron width of the \(J/\psi\) meson and the
probability of its decay to the \(\mu^+\mu^-\) pair, \GBmumu.  Given
independent data on the branching fraction 
\(\mathcal{B}_{ee}\)~\cite{PDG-2008},
we use this result to evaluate the leptonic \(\Gamma_{\ell\ell}\) and
total \(\Gamma\) widths.

\section{VEPP-4M collider and KEDR detector}
\label{sec:VEPP}

The VEPP-4M collider~\cite{Anashin:1998sj} can operate in the broad
range of beam energies from 1 to 6 GeV. The peak luminosity in the
\(J/\psi\) energy range is
about~\(2\times10^{30}\,\text{cm}^{-2}\text{s}^{-1}\).

\begin{figure}[t]
  \centering\includegraphics[width=\columnwidth]{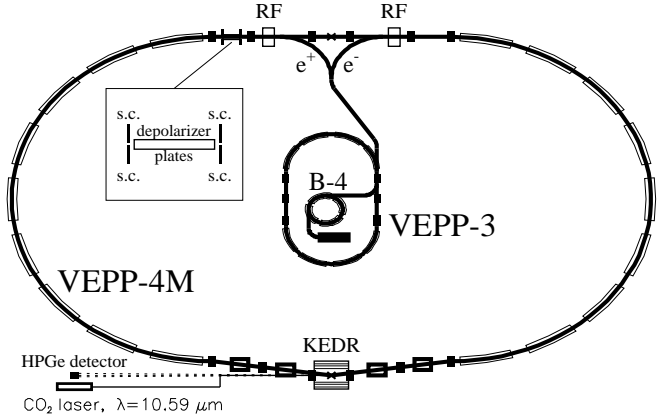}
  \caption{VEPP-4M/KEDR complex with the resonant depolarization and
    the infrared light Compton backscattering facilities.}
  \label{fig:vepp4m}
\end{figure}

One of the main features of the VEPP-4M is a possibility of precise
energy determination. The resonant depolarization
method~\cite{Bukin:1975db,Skrinsky:1989ie} was implemented at VEPP-4
at the very beginning of experiments in early eighties for the
measurements of the \(J/\psi\) and \(\psi(2S)\) mass with the
OLYA~\cite{Artamonov:2000cz} detector and \(\Upsilon\) family mass
with the MD-1~\cite{Artamonov:2000cz} detector.

At VEPP-4M the accuracy of  energy calibration with the
resonant depolarization is improved to about \(10^{-6}\). 
Between calibrations the energy interpolation
in the \(J/\psi\) energy range has  the accuracy of  
\(6\cdot10^{-6}\) ($\simeq$10~keV)~\cite{Aulchenko:2003qq}.

To monitor beam energy during data taking
the infrared light Compton
backscattering is employed (with 50$\div$70~keV precision in the
\(J/\psi\) region), which was first developed at the BESSY-I and BESSY-II
synchrotron radiation sources~\cite{Klein:1997wq,Klein:2002ky}.

The KEDR detector~\cite{Anashin:2002uj} includes the vertex detector,
the drift chamber, the scintillation time-of-flight counters, the
aerogel Cherenkov counters, the
barrel liquid krypton calorimeter, the endcap CsI calorimeter, and the
muon system built in the yoke of a superconducting coil generating a
field of 0.65 T. The detector also includes a tagging system to
detect scattered electrons and study two-photon processes. The
on-line luminosity is measured by two independent
single bremsstrahlung monitors.

\section{Experiment description}
\label{sec:Exp}

A data sample used for this analysis comprises 230 nb\(^{-1}\)
collected at 11 energy points in the \(J/\psi\) energy range.  This
corresponds to approximately 15000 \(J/\psi\to e^+e^-\) decays. During
this scan, 26 calibrations of the beam energy have been done using
resonant depolarization.

The primary trigger signal was provided by a coincidence of two
non-adjacent scintillation counters or an energy deposition in the
endcap calorimeter of at least 100 MeV.  A veto from the
endcap-calorimeter crystals closest to the beam line was used to
suppress the machine background.

The secondary trigger required at least two tracks in the drift
chamber or at least one track and an energy deposition in the
calorimeter of at least 70 MeV and the coincidence of two
non-adjacent scintillation counters.

The hardware triggers use the analogous output of the calorimeter with
reduced energy resolution.
During the offline analysis real and simulated events pass through 
the software event filter which recalculates the
trigger decision using a digitized response of the detector subsystems.
The calorimeter energy thresholds in the event filter 
are toughened by a factor of 1.5 with respect to the instrumental
values, suppressing the uncertainty in the latter and their possible
instability.

Single bremsstrahlung and \(e^+e^-\to e^+e^-\) events 
at polar angles  in the
range between \(18^{\circ}\) and \(31^{\circ}\) (the endcap calorimeter) 
were used in the relative measurement of luminosity. In
order to evaluate \GBee, it was unnecessary to measure the absolute
luminosity. Since \(e^+e^-\to e^+e^-\) events analyzed here include 
both events from the resonance
and a well known non-resonant QED background, it is
possible to perform an absolute calibration of the luminosity along
with the derivation of \GBee.

\begin{figure}[t]
  \centering\includegraphics[width=\columnwidth]{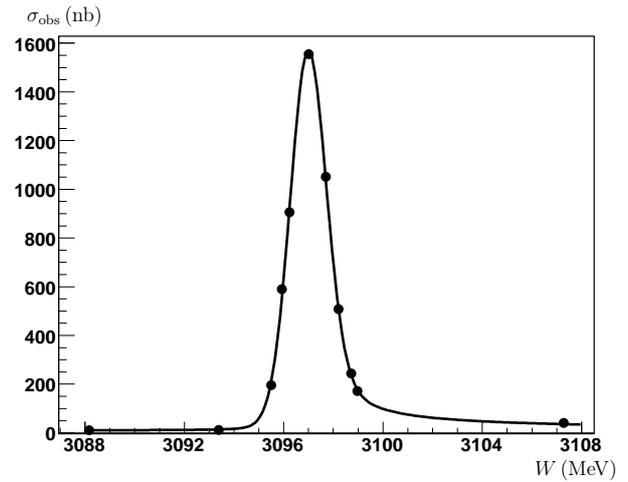}
  \caption{Observed cross section of  \(e^+e^-\to\text{hadrons}\) 
    in the \(J/\psi\) scan.}
  \label{fig:mhadr}
\end{figure}

Figure~\ref{fig:mhadr} shows the observed cross sections of 
\(e^+e^-\to\text{hadrons}\)
in the \(J/\psi\) energy range.  These data were used
to fix the resonance peak position and to determine the beam energy
spread.  The value of the \(J/\psi\) mass agrees with the earlier
VEPP-4M/KEDR experiments~\cite{Aulchenko:2003qq}. The accuracy of
the energy spread was about 2\%, including variations
associated with the beam current.

\section{Theoretical $e^+e^-\to\ell^+\ell^-$ cross section}
\label{sec:theory-simulation}

The analytical expressions for the cross section of
the process $e^+e^-\to\ell^+\ell^-$ 
with radiative corrections taken into account in the soft photon 
approximation were first derived by  Ya.\,A.~Azimov et al. 
in 1975~\cite{azimov-1975-eng}.
With some up-today modifications one obtains
in the vicinity of a narrow resonance
\begin{equation}
  \label{eq:ee2mumu}
  \begin{aligned}
    &\left(\frac{d\sigma}{d\Omega}\right)^{ee\to\mu\mu}\approx
    \left(\frac{d\sigma}{d\Omega}\right)_{\text{QED}}^{ee\to\mu\mu}+
    \frac{3}{4M^2}
    \left(1+\delta_{\text{sf}}\right)
    \left(1+\cos^2\theta\right) 
    \,\times\\
    &\quad\qquad\left\{
        \frac{3\Gamma_{ee}\Gamma_{\mu\mu}}{\Gamma M}              
        \Im \mathcal{F}- 
        \frac{2\alpha\sqrt{\Gamma_{ee}\Gamma_{\mu\mu}}}{M}\,
        \Re \frac{\mathcal{F}}{1-\Pi_0}
     \right\} ,
  \end{aligned}
\end{equation}
where a correction $\delta_{\text{sf}}$ follows from the 
structure function approach
of~\cite{KuraevFadin}:
\begin{equation}\label{eq:deltasf}
  \delta_{\text{sf}}=\frac{3}{4}\beta+
   \frac{\alpha}{\pi}\left(\frac{\pi^2}{3}-\frac{1}{2}\right)+
  \beta^2\left(\frac{37}{96}-\frac{\pi^2}{12}-
  \frac{1}{36}\ln\frac{W}{m_e} \right)
\end{equation}
and 
\begin{equation}  \label{eq:F}
  \mathcal{F}=\frac{\pi\beta}{\sin\pi\beta}\,
      \left(\frac{M/2}{-W+M-i\Gamma/2}\right)^{1-\beta}
\end{equation}
with 
\begin{equation}
\label{eq:beta}
\beta=\frac{4\alpha}{\pi}\left(\ln\frac{W}{m_e}-\frac{1}{2}\right).
\end{equation}
Here $W$ is the center-of-mass energy and $\Pi_0$ represents the
vacuum polarization operator with the resonance
contribution excluded. The terms proportional to $\Im\mathcal{F}$ and
$\Re\mathcal{F}$ describe the contribution of the resonance
and the interference effect, respectively. 
The definition of leptonic width in
Eq.~\eqref{eq:ee2mumu}--~\eqref{eq:beta}
implicitly  includes vacuum polarization 
as  recommended by PDG:
$\Gamma_{\ell\ell}=\Gamma^{0}_{\ell\ell}/|1-\Pi_0|^2$, where $\Gamma^{0}_{ee}$
is the lowest-order QED value.

The function $\mathcal{F}$ in eq.~\eqref{eq:F} 
 appears from the integration (see ~\cite{Todyshev}, where one can
also find  the definition
of $\mathcal{F}$ for the relativistic Breit-Wigner amplitude)
and differs from that in
Ref.~\cite{azimov-1975-eng} by the $\pi\beta/\!\sin\pi\beta$ factor.  


For the $e^{+}e^{-}$ final state one has
\begin{equation}
  \label{eq:ee2ee}
  \begin{aligned}
    &\left(\frac{d\sigma}{d\Omega}\right)^{ee\to ee} \approx 
    \left(\frac{d\sigma}{d\Omega}\right)_{\text{QED}}^{ee\to ee}+\\
    &\quad\frac{1}{M^2}
    \left\{\,\frac{9}{4}\frac{\Gamma^2_{ee}}{\Gamma
      M}(1+\cos^2\theta) \,\left(1+\delta_{\text{sf}}\right)\,\Im\mathcal{F} -\right.\\
    &\qquad\left.\frac{3\alpha}{2}\frac{\Gamma_{ee}}{M}
    \left [(1+\cos^2\theta)-
      \frac{(1+\cos\theta)^2}{(1-\cos\theta)}\right ]
              \Re \mathcal{F}
    \right\}.
  \end{aligned}
\end{equation}

The goal of this analysis is a measurement of \GBee and \GBmumu
contained in the resonant terms.  The precision of these
terms
in formulae~\eqref{eq:ee2mumu} and~\eqref{eq:ee2ee}
is better than 0.2\%. It was estimated
by numerical calculations beyond the soft photon
approximation according to Ref.~\cite{KuraevFadin}.
Although the interference terms could allow  a direct measurement of
\(\Gamma_{ee}\) ($e^+e^-\to e^+e^-$) and
\(\sqrt{\Gamma_{ee}\Gamma_{\mu\mu}}\) ($e^+e^-\to\mu^+\mu^-$),
in our case we are limited by  the statistical accuracy and
theoretical uncertainty.

To compare experimental data with the theoretical cross
sections~\eqref{eq:ee2mumu} and~\eqref{eq:ee2ee},
it is necessary to perform their convolution with a distribution of the total
beam energy which is
assumed to be Gaussian with an energy spread \(\sigma_W\):
\[\rho(W)=\frac{1}{\sqrt{2\pi}\,\sigma_W}\exp{\left(-\frac{(W-W_0)^2}{2\sigma_W^2}\right)}\,,\]
where \(W_0\) is an average c.m. collision energy. 

Since the energy spread \(\sigma_W\simeq0.7\,\text{MeV}\)
is much larger than the intrinsic width of the  \(J/\psi\) meson, 
the uncertainty of the cross section due to the
knowledge of the latter is suppressed. 
We use the value
\(\Gamma\simeq0.093\,\text{MeV}\)~\cite{PDG-2008}.
  
For simulating the nonresonant contribution \(\sigma_{\text{QED}}\)  
we use the 
calculations of~\cite{Beenakker:1990mb,arbuzov-1997-9710} as 
implemented in two independent generators BHWIDE~\cite{BHWIDE}
and MCGPJ~\cite{MCGPJ}. 

The resonant and interference cross sections were simulated using
simple generators with proper angular distributions.
In this case the initial state radiative corrections are already
taken into account by the expressions~\eqref{eq:ee2mumu} and \eqref{eq:ee2ee}. 
These formulae implicitly involve the branching ratios
$\Gamma_{\ell\ell}/\Gamma = \mathcal{B}_{\ell\ell(n\gamma)}$
with the arbitrary number of soft photons emitted. Actual event
selection criteria can not
be 100\% efficient for events with additional photons,
therefore the final state radiation
must be simulated explicitly. This was done using the PHOTOS
package~\cite{photos}.

\section{Data analysis}
\label{sec:Data}

In our analysis we employed the simplest selection criteria that
ensured  sufficient suppression of multihadron events and the
cosmic-ray background.  The following requirements were imposed for
\(e^+e^-\to e^+ e^-\) events selection:
\begin{enumerate}
\item An event should have exactly two oppositely charged tracks, each
  originating from the beam intersection region, having a continuation in
  the calorimeter, and lying in the range of angles between the
  particle and beam axis from \(30^{\circ}\) to \(150^\circ\).\label{item:ee:1}
\item The energy deposited in the calorimeter for each particle should
  be higher than 0.7\,GeV, and the sum of the energies of the two
  particles should be higher than 2\,GeV.
\item The energy deposited in the calorimeter and not associated with
  the two particles considered should not exceed 5\% of the
  total energy deposition.
\item The angle between selected particles
  should be larger than \(140^{\circ}\) and acoplanarity less
  than \(40^{\circ}\).
\end{enumerate}
Requirements for selecting \(e^+e^-\to \mu^+ \mu^-\) events
are:
\begin{enumerate}
\item The same as tracking criteria  for \(e^+e^-\to e^+e^-\).
\item The energy deposited in the calorimeter for each particle should
  be higher than 60\,MeV and less than 500\,MeV, and the sum of the
  energies of the two particles should not be higher than 750\,MeV.
\item The energy deposited in the calorimeter and not associated with
  the two particles considered should not exceed 30\% of the
  total energy deposition.
\item The angle between selected particles
  should be larger than \(170^{\circ}\) and acoplanarity less
  than \(15^{\circ}\).
\item The momentum for each particle should be higher than 500\,MeV$/c$, and
  the sum of the momenta of the two particles should be higher than
  2\,GeV$/c$.
\item There is at least one time measurement in the time-of-flight
  system. The uncorrected measured time should be within the
  \mbox{\([-3.75\div10.]\)}~ns  range from the beam intersection time.
\end{enumerate}

\begin{figure}[t]
  \centering\includegraphics[width=\columnwidth]{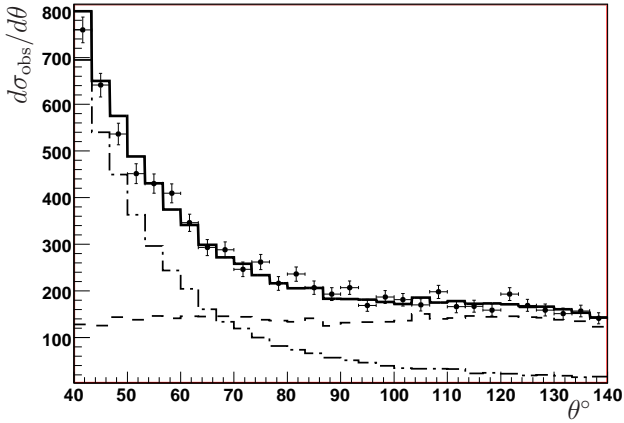}
  \caption{Cross section of the process $e^+e^-\to e^+e^-$  
    as a function of the electron scattering angle at the  $J/\psi$ peak. The points
    represent experimental data.  The histograms correspond to a
    simulation: the dashed line represents the contribution of the 
    $J/\psi$ resonance,the dashed and dotted line represents 
    the contribution of
    Bhabha scattering and the
    solid-line histogram is the sum of the first two.}
  \label{fig:a-theta}
\end{figure}

Figure~\ref{fig:a-theta} shows the distribution of selected
$e^+e^-\to e^+e^-$ events with respect to the electron scattering
angle. The displayed points represent the experimental values, while
the histograms correspond to the simulation.  At small angles Bhabha
scattering prevails, while at large angles events of resonance decay  are
dominant.  The interference effect is not shown since the
presented data correspond to the \(J/\psi\) peak, where the
interference vanishes.

In order to measure the resonance parameters, the set of events was
divided into ten equal angular intervals from \(40^{\circ}\) to
\(140^{\circ}\).  At the $i$-th   energy point \(E_i\) and the
$j$-th angular interval \(\theta_j\), the expected number of
$e^+e^-\to e^+e^-$ events was parameterized as
\begin{equation}\label{eq:real2sim}
  \begin{aligned}
    N_{\text{exp}}(E_i,\theta_j)=&\mathcal{R}_{\mathcal{L}}\times \mathcal{L}(E_i)\times\\
    \Big(&\sigma^{\text{theor}}_{\text{res}}(E_i,\theta_j)\cdot
    \varepsilon^{\text{sim}}_{\text{res}}(E_i,\theta_j)+ \\
    &\sigma^{\text{theor}}_{\text{inter}}(E_i,\theta_j)\cdot
    \varepsilon^{\text{sim}}_{\text{inter}}(E_i,\theta_j)+\\
    &\sigma^{\text{sim}}_{\text{Bhabha}}(E_i,\theta_j)\cdot
    \varepsilon^{\text{sim}}_{\text{Bhabha}}(E_i,\theta_j)
    \Big).
  \end{aligned}
\end{equation}
where \(\mathcal{L}(E_i)\) is the integrated luminosity   measured by
the luminosity monitor at
the $i$-th energy point; \(\sigma^{\text{theor}}_{\text{res}}\),
\(\sigma^{\text{theor}}_{\text{inter}}\) and
\(\sigma^{\text{theor}}_{\text{Bhabha}}\) are the theoretical
cross sections for resonance, interference and Bhabha contributions,
respectively. 
\(\varepsilon^{\text{sim}}_{\text{res}}\),
\(\varepsilon^{\text{sim}}_{\text{inter}}\) and
\(\varepsilon^{\text{sim}}_{\text{Bhabha}}\) are detector efficiencies
 obtained from simulation. The efficiencies differ mainly
 due to difference in radiative corrections.
 Unlike the Bhabha process, the initial state radiation for
 the narrow resonance production is strongly suppressed, thus the
 events are more collinear.

\begin{figure}[t]
  \centering\includegraphics[width=\columnwidth]{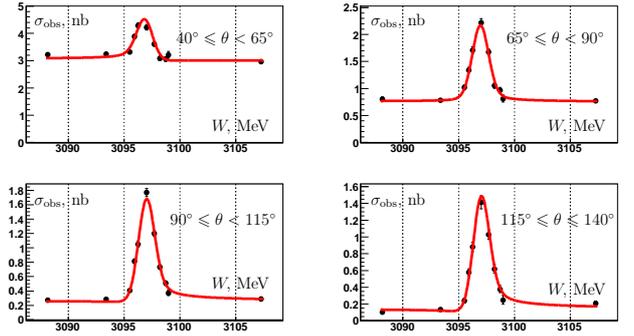}
  \caption{Fits to experimental data for the process $e^+e^-\to e^+e^-$ 
    in the $J/\psi$ energy range for four angular ranges.}
  \label{fig:real2sim}
\end{figure}
 
 In this formula the following free parameters were used:
\begin{enumerate}
\item the product \GBee, which determines the magnitude of the
  resonance signal;
\item the electron width \(\Gamma_{ee}\), which specifies the
  amplitude of the interference wave;
\item the coefficient \(\mathcal{R}_{\mathcal{L}}\), which provides
  the absolute calibration of the luminosity monitor.
\end{enumerate}

We note that the coefficient \(\mathcal{R}_{\mathcal{L}}\) partially 
takes into account a possible difference between the actual 
detection efficiency
and simulation in the case where this difference does not depend on the
scattering angle or the beam energy (or the data taking time),
thus a substantial cancellation of errors occurs.

 The \(\Gamma_{ee}\) value
obtained from the fit to the data has large statistical and systematic
uncertainties caused by the smallness of the interference effect
and the low accuracy of theoretical evaluation.

Figure~\ref{fig:real2sim} shows our fits to the data for four angular
bins.  For this fit 
\(\chi^2/\text{ndf}=53.7/41\)  taking into account only the
statistical errors and \(\chi^2/\text{ndf}\simeq40.5/41\) after
converting the energy determination uncertainty to the
cross section error.

The joint fit in ten equal bins from \(40^\circ\) to
\(140^\circ\) produces the following
basic result:
\begin{equation}\label{eq:eeresult}
  \begin{split}
    &\GBee = 0.3323\pm0.0064\,\text{(stat.)}\,\text{keV},\\
    &\mathcal{R}_{\mathcal{L}}=93.4\pm0.7\,\text{(stat.)}\,\%,\\
    &\Gamma_{ee} = 5.7 \pm0.6\,\text{(stat.)}\,\text{keV}.
 \end{split}
\end{equation}

Due to different angular distributions for Bhabha scattering and
resonance events, subdivision of the data into several angular bins reduces
the statistical error for \GBee by \(40\div50\,\%\). 
Here  \(\Gamma_{ee}\) has a statistical error of about 10\% and
agrees with the world average value. The same value can be obtained
with a much higher precision using \GBll and an independent
measurement of the branching ratio $J/\psi\to\ell^+\ell^-$.

Similarly to~\eqref{eq:real2sim}, the expected number of $e^+e^-\to
\mu^+\mu^-$ events was parameterized in the form:
\begin{equation}
  \begin{aligned}
    N_{\text{exp}}(E_i)=&\mathcal{R}_{\mathcal{L}}\times\mathcal{L}(E_i)\times\\
    \Big(&\sigma^{\text{theor}}_{\text{res}}(E_i)\cdot\varepsilon^{\text{sim}}_{\text{res}}(E_i)+\\
    &\sigma^{\text{theor}}_{\text{inter}}(E_i)\cdot\varepsilon^{\text{sim}}_{\text{inter}}(E_i)+\\
    &\sigma^{\text{theor}}_{\text{bg}}(E_i)\cdot\varepsilon^{\text{sim}}_{\text{bg}}(E_i)
    \Big)+F_{\text{cosmic}}\times T_i,
  \end{aligned}\label{eq:mumufit}
\end{equation}
with the same meaning of \(\mathcal{R}_{\mathcal{L}}\) and
\(\mathcal{L}(E_i)\) as  in~\eqref{eq:real2sim}. 
\(\mathcal{L}(E_i)\)  is multiplied
by the sum of the products of theoretical cross sections for
resonance, interference and QED background and detection efficiencies
as obtained from simulated data.  
\(\mathcal{R}_{\mathcal{L}}\) 
was fixed from the result~\eqref{eq:eeresult} and 
  \(T_i\)
is the live data taking time. Unlike~\eqref{eq:real2sim}, there is 
only one angular bin  from \(40^{\circ}\) to \(140^{\circ}\).

The following free parameters were used:  
\begin{enumerate}
\item the product \GBmumu, which determines the magnitude of the
  resonance signal;
\item the square root of electron and muon widths
  \(\sqrt{\Gamma_{ee}\Gamma_{\mu\mu}}\), which specifies the amplitude
  of the interference wave;
\item the rate of cosmic events,  \(F_{\text{cosmic}}\), that passed
  the selection criteria for the \(e^+e^-\to\mu^+\mu^-\) events.
\end{enumerate}
Due to variations of luminosity during the experiment it is possible
to separate the contribution of cosmic events  (\(F_{\text{cosmic}}\cdot T_i\))
from that of the nonresonant background 
(\(\sigma^{\text{theor}}_{\text{bg}}(E_i)\cdot\varepsilon^{\text{sim}}_{\text{bg}}(E_i)\cdot \mathcal{L}(E_i)\)).

\begin{figure}[t]
  \centering\includegraphics[width=\columnwidth]{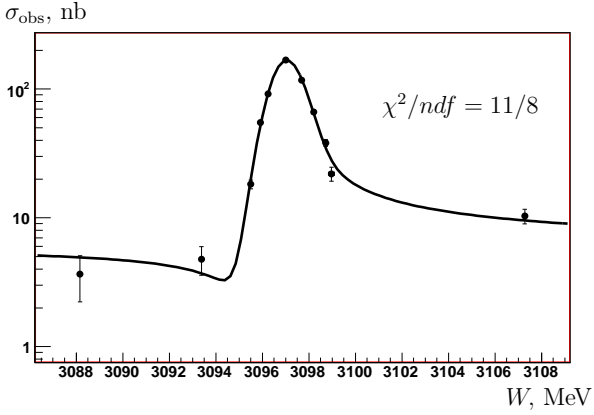}
  \caption{Fit to experimental data for $e^+e^-\to \mu^+\mu^-$ process
    in the $J/\psi$ energy range.}
  \label{fig:mumufit}
\end{figure}

Figure~\ref{fig:mumufit} shows our fit to the \(e^+e^-\to\mu^+\mu^-\)
data.  It yields the following  result:
\begin{equation}\label{eq:mmresult}
  \begin{split}
    &\GBmumu = 0.3318\pm0.0052\,\text{(stat.)}\,\text{keV},\\
    &\sqrt{\Gamma_{ee}\times\Gamma_{\mu\mu}}=5.6\pm0.7\,\text{(stat.)}\,\text{keV}.
  \end{split}
\end{equation}
As can be seen from~\eqref{eq:mmresult}, the statistical error of 
\GBmumu is about 1.6\%.

\section{Discussion of systematic uncertainties}

\label{sec:ErrDisc}

The most significant systematic uncertainties in the \GBee and 
\GBmumu measurements are
listed in Tables~\ref{tab:ee:systematic} and~\ref{tab:mumu:systematic},
respectively.
A few dominant sources of uncertainty
are briefly described below.

A rather large  uncertainty of 0.8\% common for the electron and muon
channels is due to the luminosity monitor instability.
It was estimated from comparing the results obtained using the
on-line luminosity of the single bremsstrahlung monitor and
the off-line luminosity
measured by the $e^+e^-$ scattering in the endcap
calorimeter. 

\begin{table}[t]
\centering
\caption{Systematic uncertainties in \GBee.}
\label{tab:ee:systematic}
\medskip
\begin{tabular}{lc}\hline
Systematic uncertainty source & Error,\,\%\\\hline
Luminosity monitor instability & 0.8\\
Offline event selection & 0.7\\
Trigger efficiency      & 0.5\\
Energy spread accuracy &0.2\\
Beam energy measurement (10--30\,keV) & 0.3\\
Fiducial volume cut &0.2\\
Calculation of radiative corrections & 0.2\\
Cross section for Bhabha  (MC generators) & 0.4\\
Final state radiation (PHOTOS)  & 0.4\\
Background from  \(J/\psi\) decays&  0.2\\
Fitting procedure & 0.2\\
\hline
\textit{Total}&\textit{1.4}\\\hline
\end{tabular}
\end{table}

\begin{table}[t]
\centering
\caption{Systematic uncertainties in \GBmumu.}
\label{tab:mumu:systematic}
\medskip
\begin{tabular}{lc}\hline
Systematic uncertainty source & Error,\,\%\\\hline
Luminosity monitor instability  & 0.8\\
Absolute luminosity calibration by $e^+e^-$ data &1.2\\
Trigger efficiency      & 0.5\\
Energy spread accuracy &0.4\\
Beam energy measurement (10--30\,keV) & 0.5\\
Fiducial volume cut &0.2\\
Calculation of radiative corrections & 0.2\\
Final state radiation (PHOTOS)  & 0.5\\
Nonresonant background  & 0.1\\
Background from  \(J/\psi\) decays &  0.6\\
\hline
\textit{Total}&\textit{1.9}\\\hline
\end{tabular}
\end{table}

The essential source of uncertainty is an imperfection of the detector
response simulation resulting in the errors in 
the trigger and  offline event selection efficiencies.

To correct the offline event selection efficiency, two high-purity control samples
of $e^{+}e^{-}$ events were prepared.
The first sample selected using the LKr-calorimeter data only
was employed to determine the tracking system efficiency,
the second sample obtained using mostly the tracking system data allows one
to check calorimeter related cuts. Each sample contains about
70\% of all events used in the analysis. The same analysis was performed
with simulated data.
The corrections already taken into account
in~\eqref{eq:eeresult} were
\begin{equation}\label{eq:eecorr}
  \begin{split}
    &\delta\,\GBee = 0.8\pm0.6\,\text{(stat.)}\,\pm 0.4\,\text{(syst.)}\,\%,\\
    &\delta\,\mathcal{R}_{\mathcal{L}}=1.7\pm0.5\,\text{(stat.)}\,
      \pm 0.5\,\text{(syst.)}\,\%.\\
 \end{split}
\end{equation}
The statistical error of the efficiency determination is approximately
three times less than that of the final result due to the binomial distribution
in the number of lost events. The residual systematic error is due 
to an incomplete
event sample employed for the correction and the efficiency difference
for the resonance decays and the continuum events.
The variation of  $\mathcal{R}_{\mathcal{L}}$ is greater than 
that of the main result illustrating the cancellation
of uncertainties mentioned in Sec.~\ref{sec:Data}.

Three contributions dominate the trigger efficiency uncertainty.
The inefficiency of the time-of-flight counters used in the 
first level trigger
was studied using the cosmic ray events and equals 0.3\%. 
The second contribution comes from the cut on the number of the vertex detector
tubes hit in the event. It was used in the software trigger level
for the machine background suppression. Some fraction of events was accepted
unconditionally to check the cut.
The third contribution is due to the veto from
the CsI crystals nearest to the beam line. It is negligible for
the resonance decays and reaches 0.4\% for continuum events for
which the initial state radiation is not suppressed. The quoted value
was obtained varying the threshold in the event filter
within its uncertainty. The accidental signal-background coincidences
were taken into account by the veto rate with much better accuracy.

The uncertainty of the theoretical Bhabha cross section
was estimated comparing the results obtained with
the BHWIDE~\cite{BHWIDE} and MCGPJ~\cite{MCGPJ} event generators.
It agrees with the accuracies of the generators quoted by the authors.


The dominant uncertainty of the \GBmumu result is
associated with the absolute luminosity calibration done
in the $e^+e^-$-channel.
It includes the accuracy of the Bhabha event generators,
the statistical error of $\mathcal{R}_\mathcal{L}$
from~\eqref{eq:eeresult} and the residual  efficiency difference
for $e^+e^-$ and $\mu^+\mu^-$ events after a correction using
simulated data.

To determine this residual difference and make a proper correction,
samples of real and simulated quasi-collinear events were selected 
using an  alternative track reconstruction
code finding a single track with a kink at the point of the closest
approach to the beam line. Then the standard analysis procedure
was performed for these samples yielding  the double ratio
\begin{equation}\label{eq:mmcorr}
 \left( \frac{\epsilon_{\mu\mu}^{\text{exp}}}{\epsilon_{\mu\mu}^{\text{sim}}} \right) \Big/
   \left( \frac{\epsilon_{ee}^{\text{exp}}}{\epsilon_{ee}^{\text{sim}}} \right) =
 1.005 \pm0.005\,\text{(stat.)}\,\pm 0.008\,\text{(syst.)}.
\end{equation}
The sample selected contains about 80\% and 50\% of all $\mu^{+}\mu^{-}$ 
and $e^{+}e^{-}$ events, respectively. The systematic error of the
ratio reflects the incompleteness of the samples.

The trigger veto uncertainty is the same as for \(J/\psi\to e^+e^-\) decay.

The background for   \(J/\psi\to\mu^+\mu^-\) decay from
hadronic decays of \(J/\psi\)
was estimated with the help of the
muon system. It contributes \(1.5\pm0.6\,\%\) to the
selected \(\mu^+\mu^-\) events. The estimation agrees with the simulation
results.
This correction as well as the correction~\eqref{eq:mmcorr} 
have already been taken into account
in the \GBmumu result~\eqref{eq:mmresult}.

Due to the high precision in the
 energy determination by the
resonant depolarization method~\cite{Anashin:2007zz}, the corresponding
errors (peak position, energy spread, and energy
at a point) are relatively small.

In calculating the cross section for resonance production with
formulae~\eqref{eq:ee2mumu} and~\eqref{eq:ee2ee} we used the PDG
value of the total width \(\Gamma\) from~\cite{PDG-2008}. Its error 
is about 2\%, which gives a $\sim 0.05\%$ contribution to the 
error in our result.

The fiducial volume cut $40^{\circ} < \theta < 140^{\circ}$ was applied
using the tracking system and the strip system of the LKr calorimeter.
The difference of results provides a conservative uncertainty
estimate.

All other uncertainties are rather clear. More detail can be found 
in~\cite{baldin-preprint-09-eng}.

All the uncertainties for \GBee added in quadrature 
yield a systematic error of 1.4\%. All uncertainties for \GBmumu
added in quadrature  yield a systematic error of 1.9\%.

\section{Results and Conclusion}

\begin{figure}[t]
  \centering\includegraphics[width=\columnwidth]{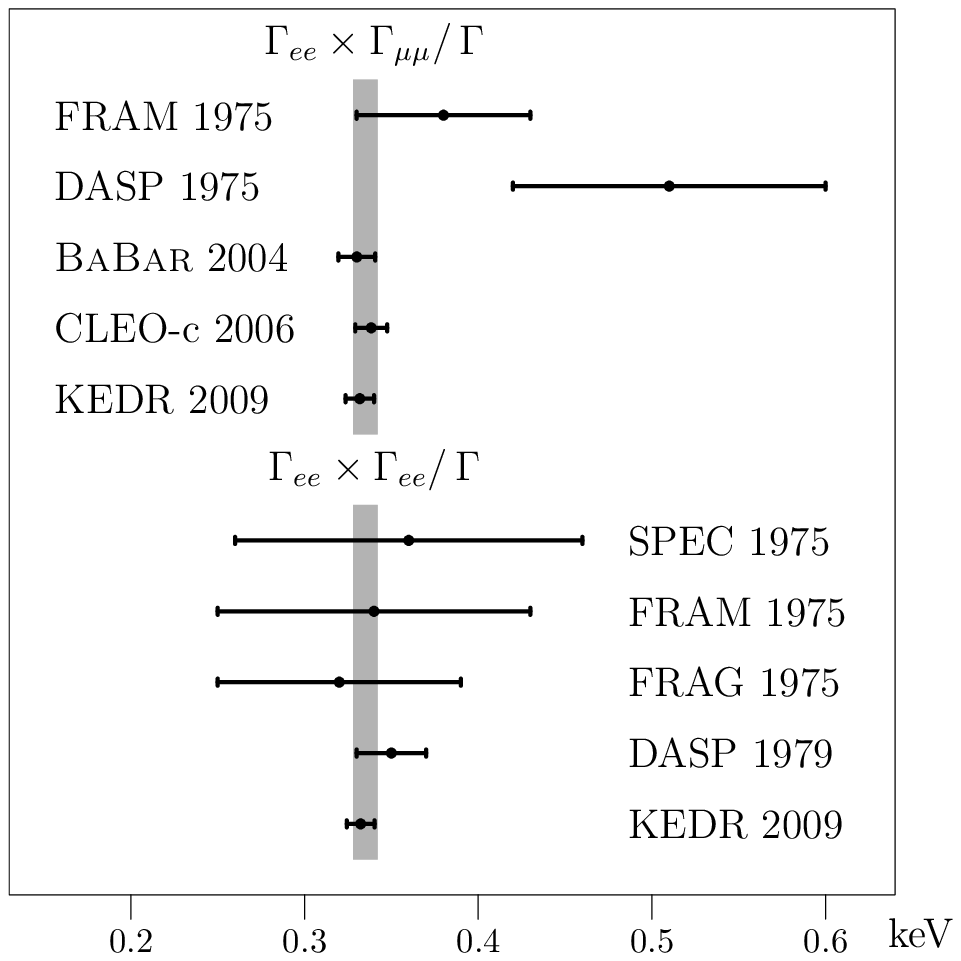}
  \caption{Comparison of \GBee and \GBmumu measured in different
    experiments mentioned in~\cite{PDG-2008} with KEDR 2009 results.
    The grey strip is for the world average \GBmumu value.}
  \label{fig:cchist}
\end{figure}

The new measurement of the \GBee and \GBmumu has been performed at the
VEPP-4M collider using the KEDR detector.  The following results have
been obtained:
\begin{equation*}
  \begin{split}
    &\GBee\,=0.3323\pm0.0064\,\text{(stat.)}\,\pm0.0048\,\text{(syst.)}\,\,\text{keV,}\\ 
    &\GBmumu=0.3318\pm0.0052\,\text{(stat.)}\,\pm0.0063\,\text{(syst.)}\,\,\text{keV.}   
  \end{split}
\end{equation*}

Previously, \GBee was measured in the DASP experiment in
1979~\cite{Brandelik:1979hy} with a precision of about 6\%. The result
obtained in the present study improves the accuracy by
a factor greater than two. The most precise previous measurements 
of \GBmumu were
made in the \textsc{BaBar}~\cite{collaboration-2004-69} and
CLEO-c~\cite{adams-2006-73} experiments, both with the ISR technique.

Figure~\ref{fig:cchist} shows the comparison of our results 
with those of the
previous experiments.
The grey line shows the PDG average and the error for the \GBmumu product
measurement. The new KEDR results are the most precise. Results are in
good agreement with each other and with the world average value of
\GBmumu.


From the direct measurements of the products above one can
extract the leptonic and full width of the resonance as well as 
test leptonic universality. For the former one should calculate 
the sum of \GBee and
\GBmumu, while for the latter the ratio of these quantities can be used. 

While estimating uncertainties of
\(\Gamma_{ee}\times(\Gamma_{ee}+\Gamma_{\mu\mu})\,\Gamma\) and \(\Gamma_{ee}/\,\Gamma_{\mu\mu}\)
 correlations
between \GBee and \GBmumu systematic errors were taken into account:
\begin{equation*}
  \begin{split}
    \Gamma_{ee}\times(\Gamma_{ee}+\Gamma_{\mu\mu})/\,\Gamma&=
    0.6641\pm0.0082\,\text{(stat.)}\,\pm0.0100\,\text{(syst.)}\,\text{keV,} \\
    \Gamma_{ee}/\,\Gamma_{\mu\mu}&=1.002\pm0.021\,\text{(stat.)}\,\pm0.013\,\text{(syst.)} \\
  \end{split}
\end{equation*}
In contrast to the \GBee and \GBmumu values, the ratio
\(\Gamma_{ee}/\,\Gamma_{\mu\mu}\) is not sensitive to the absolute
luminosity calibration. Therefore, the $\mathcal{R}_{\mathcal{L}}$
parameter has been fixed in the fit and the relative
statistical uncertainty of the \(\Gamma_{ee}/\,\Gamma_{\mu\mu}\) value
is less than that of
$\Gamma_{ee}\times(\Gamma_{ee}+\Gamma_{\mu\mu})/\,\Gamma$.

With the assumption of leptonic universality and using independent data
on the branching fraction \(\mathcal{B}(J/\psi\to
e^+e^-)=(5.94\pm0.06)\,\%\)~\cite{PDG-2008}, the leptonic and total
widths of the \(J/\psi\) meson were determined:
\begin{equation*}
  \label{eq:result:Gll}
  \begin{split}
    &\Gamma_{\ell\ell}=5.59\pm0.12\,\,\text{keV}, \\ 
    &\Gamma\,=\,94.1\pm2.7\,\,\text{keV}. 
  \end{split}
\end{equation*}
These results are in good agreement with the world average~\cite{PDG-2008}
and with the results from
the \textsc{BaBar}~\cite{collaboration-2004-69} and
\mbox{CLEO-c}~\cite{adams-2006-73} experiments.

\section*{Acknowledgments}

We greatly appreciate the efforts of the staff of VEPP-4M to provide
good operation of the complex and the staff of experimental
laboratories for the permanent support in preparing and performing
this experiment. The authors are grateful to E.\,A.~Kuraev and
V.\,S.~Fadin for a discussion of various problems related to 
theoretical cross section
 representation.

This work was partially supported by the Russian Foundation for Basic Research,
Grant 08-02-00258 and RF Presidential Grant for Sc. Sch. NSh-5655.2008.2.

\bibliographystyle{h-physrev5}
\bibliography{article-eng}

\end{document}